\documentclass[12pt,preprint]{aastex}
\begin{document}

\title{Formation and Detectability of Terrestrial Planets around $\alpha$ Centauri B}

\author{Guedes, J. M., Rivera, E. J., Davis, E., Laughlin, G.}
\affil{Department of Astronomy and Astrophysics, University of California, 1156 High Street, Santa Cruz, CA 95060}

\author{Quintana, E. V.}
\affil{SETI Institute, 515 N. Whisman Road, Mountain View, CA 94043}

\author{Fischer, D. A.}
\affil{Department of Physics and Astronomy, San Francisco State University, San Francisco, CA}

\begin{abstract}
We simulate the formation of planetary systems around $\alpha$ Centauri B.
The N-body  accretionary evolution of a $\Sigma \propto r^{-1}$ disk populated 
with 400-900 lunar-mass protoplanets is followed for 200 Myr.  All simulations lead 
to the formation of multiple-planet systems with at least one planet in the 
1-2 M$_{\oplus}$ mass range at 0.5-1.5 AU.  We examine the detectability of 
our simulated planetary systems by generating synthetic radial velocity 
observations including noise based on the radial velocity residuals to 
the recently published three planet fit to the nearby K0\textrm{V} star 
HD 69830.  Using these synthetic observations, we find that we can reliably 
detect a 1.8 M$_{\oplus}$ planet in the habitable zone of $\alpha$ Centauri B 
after only three years of high cadence observations.  We also find that the 
planet is detectable even if the radial velocity precision is 3 m\,s$^{-1}$, 
as long as the noise spectrum is white.  Our results show that the greatest 
uncertainty in our ability to detect rocky planets in the $\alpha$ Centauri 
system is the unknown magnitude of ultra-low frequency stellar noise.
\end{abstract}


\section{Introduction}

In the past decade, over 270 extrasolar planets have been discovered in a
plethora of diverse environments.  Earth-like planets in habitable-zone 
orbits, however, remain well below the threshold of detection.  A good
representation of the Doppler velocity state-of-the-art is presented by the
triple planet system orbiting HD 69830.  This system has been shown to contain
three Neptune-mass planets, including one on a 197-day orbit, all revealed
after only 74 radial velocity observations with residual noise of
0.6 m\,s$^{-1}$ \citep{lovis06}.  The detection of the HD 69830 system suggests
that focused efforts on selected stars may be able to probe down to the
characterization of planets with radial velocity half-amplitudes considerably
below 1 m\,s$^{-1}$.  Targeted planet search around nearby stars may prove 
to be an efficient and inexpensive path to detection. Simulations of 
transiting terrestrial planets around low-mass stars show that a space-based 
search can unveil these planets within a year \citep{montgomery07} .  
In this paper, our goal is to argue that the $\alpha$ Centauri system 
provides a remarkable test-bed for pushing the Doppler detection envelope.

The $\alpha$ Centauri system, with $d$=1.33 pc, is the Sun's closest
neighbor.  It is a triple star system composed of the central $\alpha$ Cen AB
binary and the M dwarf Proxima Cen, which orbits the AB pair with a semi-major
axis of over 10,000 AU \citep{jeremy06}.  The G2V star $\alpha$ Cen A and the
K1V $\alpha$ Cen B have masses similar to the Sun with
$M_A = 1.105 \pm 0.007 M_{\odot}$ and $M_B = 0.934 \pm 0.007 M_{\odot}$ 
while the M dwarf Proxima Cen is significantly smaller with
$M_C = 0.107 \pm 0.021  M_{\odot}$ \citep{pourbaix02}.  Though both A and B
have super-solar metallicities $[Fe/H]_A = 0.22 \pm 0.02$ and
$[Fe/H]_B = 0.26 \pm 0.04$ \citep{ch92}, no planets with masses comparable to 
Neptune or larger have yet been found orbiting either star.

Astronomical observations of $\alpha$ Cen A and B have been conducted for
over 150 years.  Astrometric observations date back to the first half of the 
nineteenth century, and the radial velocities of both components have been
tabulated since 1904.  \cite{pourbaix02} simultaneously fit all the
published astrometric measurements and radial velocities of both components to
constrain the binary orbital parameters.  (See their reference list for an
historical listing of publications relating to observations of $\alpha$ Cen A
and B.)  A similar study had been done by \cite{pourbaix99}.  In the
earlier study, after fitting for the binary orbit, they examined the
plausibility of planetary companions and found that they could have detected
a planetary companion with a mass above 10 $M_{\rm Jup}$.  \cite{Endl01}
performed a similar, but more sensitive analysis, using the orbit of
\cite{pourbaix99} and their own high precision ESO Coud\'{e} Echelle
Spectrometer radial velocities.  They found upper limits for planets on
circular orbits at any orbital radius around each component of 2.5 $M_{\rm Jup}$
for $\alpha$ Cen A and 3.5 $M_{\rm Jup}$ for $\alpha$ Cen B.  We performed a
procedure in which we used the parameters from \cite{pourbaix02} and their 
radial velocities and those from \cite{Endl01} for $\alpha$ Cen B.  We used 
the systemic console (Rivera et al. 2008 in prep.), to fit only the mean anomaly 
and the velocity offsets between the three data sets.  The remaining parameters 
were held fixed.  If we simply assume that the RMS of the fit, 9 m\,s$^{-1}$, 
corresponds to the upper limit of the radial velocity half amplitude of a planet 
in a circular orbit at 1 AU or at 3 AU, then the corresponding upper limit on 
the mass of the planet is $\sim$0.3 $M_{\rm Jup}$ or $\sim$0.5 $M_{\rm Jup}$, 
respectively.  This also follows from the formula for the radial velocity half 
amplitude $K$ of a planet of mass $m_{pl}$, period $P$, and eccentricity $e$ 
orbiting a star of mass $M_{\star}$ with orbital inclination relative to the 
plane of the sky $i$, 
\begin{equation}
K = \left ( \frac{2\pi G}{P} \right )^{1/3} \frac{m_{pl}\sin{i}}{(M_{\star}+m_{pl})^{2/3}} \frac{1}{\sqrt{1-e^2}} \sim  29.8 \frac{m_{pl}\sin{i}}{\sqrt{M_{\star}a}} \rm{m\,s}^{-1}, 
\end{equation}
where for the last relation, Kepler's third law is applied, $m_{pl} \ll M_{\star}$, 
$e=0$, and the units for $m_{pl}$, $M_{\star}$, and $a$ are in $M_{Jup}$, 
$M_{\odot}$, and AU, respectively.

Several studies of $\alpha$ Cen A and B show that terrestrial planet
formation is possible around both stars despite their strong binary
interaction \citep{quintana02,quin06,quin07}.  Results to date consistently
indicate that planetary systems with one or more Earth-mass planets can form
within 2.5 AU from the host stars and remain stable for gigayear scales.
Numerical simulations and stability analyzes of planetesimal disks indicate
that material is stable within 3 AU from A/B, as long as the inclination of
the disk with respect to the binary is $\lesssim 60^{\circ}$ 
\citep{quintana02,wie97}.  In essence, with regard to the formation process,
the companion star plays the perturbative role that the gas giants in our solar
system are believed to have played during the formation phases of the Sun's
terrestrial planets.  The perturbations allow for the accretion of a large
number of planetary embryos into a final configuration containing 3-4 bodies
\citep{quintana02}. Circumprimary planet formation is known to occur in binary 
systems and despite observational selection biases, $\sim20\%$ of all planets 
discovered to date belong to multiple systems \citep{egg07}.  Of the binaries 
that are known to harbor planets, three (HD 41004,  $\gamma$ Cephei, and Gl86) 
have projected semi-major axes of $\simeq$ 20 AU, an orbital separation similar 
to that of the AB pair.  

In this paper we assess the detectability of terrestrial planets around
$\alpha$ Cen B.  We begin by carrying out eight simulations of the late stage
of planet formation using the initial conditions detailed in \S ~\ref{ICs}.
A brief description of the systems we formed is given in \S ~\ref{formation}.
Each planetary system is then tested for detectability using a Monte Carlo
method for generating synthetic radial velocity observations, as described in
\S ~\ref{detection}.  Based on our results, we are able to accurately evaluate
the detectability of planetary systems around the star.  Finally, in
\S ~\ref{discussion} we summarize our work and discuss our results.

\section{Initial Conditions}\label{ICs}

The initial conditions of the circumstellar disk in our simulations mimic
conditions at the onset of the chaotic growth phase of terrestrial planet
formation \citep{kokubo98,kenyon06} in which collisions of isolated embryos, 
protoplanets of approximately lunar mass, dominate the evolution of the disk. 
During this phase, gravitational interactions among planetary embryos serve 
to form the final planetary system around the star and clear out the remaining
material in the disk.  At the start of this phase, several hundred protoplanets
orbit the star on nearly circular orbits.

We model the $\alpha$ Centauri B circumstellar disk with a 
$\Sigma = \Sigma_0 (a/1AU)^{-1}$ surface density profile where 
$\Sigma_0 = 8.4-18.8$ g cm$^{-2}$ as calculated from the total mass $M = Nm$, 
where $N$ is the number of protoplanetary embryos in the disk and $m$ is the 
embryo's mass. These surface densities are based on disks modeled by 
\cite{chambers01} and account for the enhanced metallicity of $\alpha$ Cen B 
with respect to the Sun.  The disk extends from $1 < a < 3.5$ AU and it is 
coplanar with the binary orbit.  For each run, we populate the disk with 
$N = 400$ to $N=900$ embryos of lunar mass ($m = 0.0123 M_{\oplus}$) with 
semi-major axes chosen via a rejection method in $a$ to obtain a $\Sigma 
\propto r^{-1}$ density profile. 

Initial orbital elements of each embryo are randomly generated with mean
anomalies, arguments of pericenter, and longitudes of ascending node
extending from 0$^{\circ}$ to 360$^{\circ}$, eccentricities in the range
$0 < e < 0.001$, and inclinations in the range 
$0^{\circ} < i < 1^{\circ}$ with respect to the plane of the binary.

Integrations are run using a specialized version of the symplectic hybrid
integrator in the MERCURY integration package \citep{chambers99}.
This $N$-body code is designed to study planet growth in the presence
of a binary companion \citep{chambers02}.  Bodies grow via accretion through
perfectly inelastic embryo-embryo collisions, and therefore close encounters
are integrated directly rather than symplecticaly.  Each simulation was evolved
for 200 Myr, consuming a total of $\sim$600 cpu hours on several (dual) Intel
Xeon machines with clock speeds of at least 2.2 GHz.

We focus on terrestrial planet formation around $\alpha$ Cen B, for which we
perform a total of eight integrations named rN\_n, where N is the initial 
number of protoplanets and n is an identifier.  For instance, r700\_2 corresponds 
to our second simulation of a disk initially containing 700 bodies. As shown 
in \cite{quintana02} and \cite{quin07}, planet formation around $\alpha$ Cen A 
is expected to be qualitatively similar.

\section{Terrestrial Planet Formation}\label{formation}

Our \textit{N}-body simulations take place in the wide binary regime, with
$\alpha$ Cen B as the central star and $\alpha$ Cen A orbiting with binary
semi-major axis $a_{AB}=23.4$ AU and eccentricity $e_{AB}=0.52$
\citep{pourbaix02}.  We adopt the stellar masses to be $M_A = 1.105 M_{\odot}$ 
and $M_B = 0.934 M_{\odot}$ \citep{pourbaix02} and radii $R_A = 1.224 R_{\odot}$ 
and $R_B = 0.862 R_{\odot}$ \citep{ker03}.  Due to its low mass and large 
distance from the AB pair, Proxima Cen is neglected in our simulations.

Figure 1 shows the late evolutionary stage of a protoplanetary disk
initially containing 600 moon-mass embryos (r600\_1, see Figure 2 and Table 1).  
The radius of each circle is proportional to the radius of the object.  Bodies 
in the outer parts of the disk ($a > 3$ AU) are immediately launched into highly 
eccentric orbits and either migrate inward to be accreted by inner bodies, 
collide with the central star, or are ejected from the system ($a_{ej} = 100$ 
AU).  In this simulation, $\sim 65 \%$ of the total initial mass is cleared 
within the first 70 Myr. By the end of simulation r600\_1, four planets have 
formed.  One planet has approximately the mass of Mercury and is located at 
$a=0.2$ AU, two 0.6 $M_{\oplus}$ planets form at $a=0.7$ and $a=1.8$ AU, and a 
1.8 $M_{\oplus}$ planet forms at $a=1.09$ AU.

Table 1 shows the orbital elements of the final systems that emerge
from the calculations.  All of our simulations result in the formation
of 1-4 planets with semi-major axes in the range $0.7 < a < 1.9$ AU,
in agreement with \cite{quintana02}. We find that 42 \% of all planets
formed with masses in the range 1-2 $M_{\oplus}$ reside in the star's habitable
zone (Fig. 2), taken to be $0.5 < a_{hab} < 0.9$ \citep{kas93}.  

As a general trend, we find that disks with higher initial surface densities 
are able to retain more mass (see figure 3) but do not necessarily form more 
planets (see table 1).  All of our disks form systems with one or two planets 
in the 1-2 $M_{\oplus}$ mass range.

\section{Detectability}\label{detection}

\subsection{Why $\alpha$ Cen B is the Best Radial Velocity Candidate Star in the Sky}

The radial velocity detection of Earth-mass planets near the habitable
zones of solar type stars requires cm\,s$^{-1}$ precision.  $\alpha$ Cen B is
overwhelmingly the best star in the sky for which one can contemplate mounting
a high-cadence search.

Both $\alpha$ Cen A and B have relatively high metallicities, with
$[Fe/H]_A=0.22$ for A and $[Fe/H]_B=0.26$ for B \citep{ch92} and therefore
would have been presumably endowed with circumprimary disks containing a 
relatively high fraction of solid material.  Simulations such as the ones we have 
performed and others indicate that the final mass present in terrestrial planets 
is in direct proportion to the initial amount of material available.

$\alpha$ Cen B is exceptionally quiet, both in terms of acoustic p-wave
mode oscillations and chromospheric activity.  Observations of $\alpha$ Cen A
with the UVES Echelle Spectrograph show that the star exhibits p-mode
oscillations with amplitude varying from 1-3 m\,s$^{-1}$ and a periodicity 
of $\sim$5 minutes \citep{butler04}.  UVES observations of $\alpha$ Cen B show
that peak amplitude noise for this star is much lower, reaching only
0.08 m\,s$^{-1}$ \citep{kj05}.  Furthermore, the average noise lies in the
frequency range 7.5-15 mHz for $\alpha$ Cen A and 4.1 mHz
for $\alpha$ Cen B.  These frequencies are far higher than
the $10^{-5}$ to $10^{-4}$ mHz frequency range associated with the periods
of the putative terrestrial planets.  A focused high cadence approach involving
year-round, all-night observations would effectively average out the star's
p-mode oscillations.  The long and short term chromospheric variability of the 
$\alpha$ Cen system was studied by \cite{robrade05} using X-ray data taken 
with XMM-Newton over a period of two years.  They find that $\alpha$ Cen A's 
X-ray luminosity declined by a factor of ten in this time period, an indication 
of a moderate coronal activity.  In turn, $\alpha$ Cen B's X-ray brightness 
varied only within a factor of two, denoting rather low short-term 
chromospheric activity associated with weak stellar flares. To date, no 
long-term variability has been detected in either star. 

$\alpha$ Cen B is remarkably similar in age, mass, and spectral type 
to HD 69830, the nearby K0 dwarf known to host three Neptune-mass planets
\citep{lovis06}.  Both $\alpha$ Cen B and HD 69830 are slightly less massive
than the Sun with masses 0.91 $M_{\odot}$ and 0.86 $M_{\odot}$, respectively.
Their estimated ages are 5.6-5.9 Gyr for $\alpha$ Cen B \citep{yildiz07} and
4-10 Gyr for HD 69830.  Both stars are slightly cooler than the Sun:
$\alpha$ Cen B is a K1V with $T_{eff} = 5,350$ K, while HD 69830 is a
type K0V star with $T_{eff} =  5,385$ K.  The stars have also similar visual
absolute magnitudes, $M_V$ = 5.8 for $\alpha$ Cen B and $M_V$ = 5.7 for
HD 69830 \citep{perry97}; however, due to its proximity to us, the former star
appears much brighter ($m_V = +1.34$), allowing for exposures that are
$\sim$60 times shorter.  One can thus use a far smaller-aperture telescope,
or alternatively, entertain a far higher observational cadence.
Indeed, $\alpha$ Cen B is so bright that the CCD readout will be the primary 
limiter to an observational strategy.  Furthermore, immediate proximity
to $\alpha$ Cen A provides the opportunity to create a parallel set
of observations for both stars.  Periodicities common to both stars would be
indicative of erroneous signals being introduced by the observational pipeline
itself.  An advantage of $\alpha$ Cen B over HD 69830 is its increased
metallicity ([Fe/H] = 0.26 vs. [Fe/H] = -0.05).  Higher metallicity leads
to deeper lines, which can improve the precision of Doppler velocities.
Oscillatory p-mode noise for HD 69830 was estimated to lie between 0.2 and 
0.8 m\,s$^{-1}$ \citep{lovis06}, reaching the upper limit of the p-mode noise 
expected for $\alpha$ Cen B.

Because $\alpha$ Cen B is sightly less massive than the Sun, terrestrial
planets would induce a larger radial velocity half-amplitude (Earth induces 
a 9 cm\,s$^{-1}$ reflex velocity on the Sun).  Also, $\alpha$ Cen B is
significantly less luminous than the Sun and thus its habitable zone is closer
in \citep{kas93}.  Yet another advantage is the fact that planets should be
close to circumplanar with the binary plane, which is inclined only
$11^{\circ}$ to the line of sight to Earth.  This ensures that
$\sin{i} \simeq 1$, and that the planets will contribute nearly their full
mass to the observed radial velocity half-amplitude.

Finally, the system is perfectly positioned in the southern sky.
$\alpha$ Centauri lies at $-60^{\circ}$ declination, allowing for observations 
nearly 300 days out of the year at the latitude of the Las Campanas Observatory
or the Cerro Tololo International Observatory in Chile.

All these criteria make $\alpha$ Cen B the ideal host and candidate for the
detection of a planetary system that contains one or more terrestrial planets.

\subsection{Synthetic Data}

We took the orbital elements of the systems emerging from our simulations and
generated model radial velocities.  We developed a code which effectively
simulates the observing conditions for any specific location on Earth.  Given
the latitude and longitude of an observatory and the RA and DEC of an object,
the code determines when the object is observable.  Two additional inputs
concern the beginning and end of an observing night: the angle of the sun below
the local horizon and the maximum airmass of the object, beyond which observing
should not continue.  Some fraction of nights are lost to emulate adverse
weather conditions and other effects which could result in missed observations.
For this paper we assumed a 25\% probability for the loss of a night.
We assumed access to a dedicated telescope at Las Campanas Observatory.
At this location, $\alpha$ Centauri is observable for about 10 months out of
the year.  We assumed an observing cadence of one exposure every 200 seconds,
corresponding to the read out time of the detector (this is a conservative
estimate, since in practice we would expect a considerably higher duty cycle).
Finally, we assumed various values of Gaussian white noise to add to the model
radial velocities.

Figure 4 shows a simulated set of radial velocities for the system represented 
in run r600\_1.  In this case, we assumed Gaussian white noise with amplitude 
3 m\,s$^{-1}$.  It is important to note that the model radial velocity for 
$\alpha$ Cen B due to the four terrestrial planets in this system has an 
amplitude of 23 cm\,s$^{-1}$, or a factor of 13 below the noise. Over the five 
year span, 97,260 measurements can be obtained.  For this example, we find that 
we can confidently detect 2 or 3 planets in this time span.  Most of the
currently known extrasolar planets were announced after the false alarm
probability (FAP) for seeing a peak in the periodogram of the radial velocity
data fell below $\sim$1\%.  Many were announced after the FAP fell below this
level and additional (and sometimes extensive) statistical tests of
significance were performed.  Using the 1\% FAP as an initial indicator of the
significance of a detection for the example discussed here, we can confidently 
detect at least one of the planets after about the first three years of
observations.

In the absence of significant coloration to noise, the example shown in figures
4 and 5 is a pessimistic case.  In keeping with the fact that the star is a 
virtual twin of HD 69830 \citep{lovis06}, $\alpha$ Cen B is expected to show
very little radial velocity noise or jitter.  The detection of three Neptune
mass planets around HD 69830 was facilitated by the small instrumental
uncertainties for its radial velocity observations, with a median value of
$\approx$0.7 m\,s$^{-1}$, along with the assumption of a very small long-term 
stellar jitter.  For the example shown in Figures 4 and 5, and assuming an 
optimistic level of Gaussian noise with amplitude $\approx$0.7 m\,s$^{-1}$, 
the largest peaks in the periodograms in Figure 5 would have power of order 
10 times greater.  Additionally, those peaks would have significantly smaller
FAP's, and these FAP's would clearly reach the 1\% level in less time than in
the case with a larger amount of noise.  Thus, at least the first planet 
would have been clearly detected within the first two years, and at least 3 
of the planets would have been detected after five years.

Assuming that 3 m\,s$^{-1}$ Gaussian white noise is an upper limit to the level
of radial velocity noise expected for $\alpha$ Cen B, and that small
terrestrial planets ($M_P \lesssim2\,M_{\oplus}$) orbit this star within 3 AU,
high cadence, relatively moderate term radial velocity observations should
clearly detect at least one of these planets.  Our simulations and those of
\cite{quin07} show that for disks with small inclinations relative to the
binary orbit, large terrestrial planets tend to form in or near the habitable
zone of $\alpha$ Cen B.  Thus, it is possible we may detect a habitable
terrestrial planet around at least one of our nearest stellar neighbors.

This is true for all the models explored and for an assumed Gaussian white
noise source with amplitude up to 3 m\,s$^{-1}$, which is likely much larger
than the expected jitter for $\alpha$ Cen B.

\section{Discussion}\label{discussion}
The space density of binary systems containing roughly solar-mass
components is roughly $n=0.02$ pc$^{-3}$.  We are thus remarkably lucky
that the $\alpha$ Cen system is currently only 1.33 pc away.
The possibility that detectable terrestrial planets are orbiting
$\alpha$ Cen B does much to fire the imagination, and indeed, a positive
identification of such a planet would be a truly landmark discovery.
$\alpha$ Cen's proximity allows one to envision space-based follow-up
efforts (astrometric, coronographic, interferometric) to characterize
the planets that would be far more difficult to carry out for planets
orbiting less luminous and more distant stars.

Alternately, or to phrase the situation another way, our current understanding
of the process of terrestrial planet formation strongly suggests that
{\it both} components of the $\alpha$ Cen system should have terrestrial
planets.  A lack of planets orbiting these stars would thus provide a critical
hint that there is a significant qualitative gap in our understanding
of planet formation.

The situation is thus as follows.  A successful detection of terrestrial
planets orbiting $\alpha$ Cen B can be made within a few years and with the
modest investment of resources required to mount a dedicated radial-velocity
campaign with a 1-meter class telescope and high-resolution spectrograph.
The plan requires three things to go right.  First, the terrestrial planets
need to have formed, and they need to have maintained dynamical stability
over the past 5 Gyr.  Second, the radial velocity technique needs to
be pushed (via unprecedentedly high cadence) to a degree where planets
inducing radial velocity half-amplitudes of order cm\,s$^{-1}$ can
be discerned.  Third, the parent star must have a negligible degree of
red noise on the ultra-low frequency range occupied by the terrestrial
planets.

In this paper, we've made the case that conditions 1 and 2 are highly
likely to have been met.  In our view, the intrinsic noise spectrum of
$\alpha$ Centauri B is likely all that stands between the present day and
the imminent detection of extremely nearby, potentially habitable planets.
Because whole-sun measurements of the solar noise are intrinsically
difficult to obtain, our best opportunity to measure microvariability
in radial velocities is to do the $\alpha$ Cen AB Doppler experiment.
The intrinsic luminosity of the stars, their sky location, and their
close pairing will allow for a definitive test of the limits of the
radial velocity technique.  If these limits can be pushed down to the
cm\,s$^{-1}$ level, then the prize, and the implications, may be very great
indeed.

\acknowledgements
We thank the anonymous referee for providing useful comments that improved 
our paper. This research was conducted with financial support from NASA PGG 
grant NNG04GK19G and NSF grant AST 0049986 to G.L and a NSF Graduate Research 
Fellowship to J.M.G.

\clearpage

\begin{table}
\begin{center}
\caption{Simulation Results\label{table1}}
\begin{tabular}{|c|c|c|c|c|c|c|c|c|}
\tableline
Run & $N$ & $\Sigma_0$ [ g cm$^{-2}$ ] & planet & $M$ [ $M_{\oplus}$ ] & Period [ yr ]  & $a$ [ AU ] & $e$ & $I$ [$^{\circ}$] \\
\tableline
\tableline
r900\_1 & 900 & 18.8 & a & 2.054 & 0.760 & 0.806 & 0.052 & 1.585 \\
        &     &      & b & 0.922 & 2.412 & 1.734 & 0.051 & 5.784 \\
        &     &      & c & 0.036 & 0.361 & 0.491 & 0.094 & 18.108 \\
        &     &      & d & 1.291 & 1.464 & 1.248 & 0.145 & 5.391 \\
\tableline
r800\_1 & 800 & 16.7 & a & 0.086 & 0.227 & 0.361 & 0.244 & 19.135 \\
        &     &      & b & 1.316 & 0.495 & 0.606 & 0.105 & 1.639 \\
        &     &      & c & 1.279 & 1.453 & 1.242 & 0.168 & 2.042 \\
\tableline
r800\_2 & 800 & 16.7 & a & 0.996 & 1.769 & 1.420 & 0.169 & 6.034 \\
        &     &      & b & 0.098 & 0.441 & 0.563 & 0.325 & 8.259 \\
        &     &      & c & 2.435 & 0.835 & 0.858 & 0.024 & 3.759 \\
\tableline
r700\_1 & 700 & 14.7 & a & 0.897 & 2.262 & 1.669 & 0.198 & 4.965 \\
        &     &      & b & 2.165 & 0.812 & 0.843 & 0.142 & 4.516 \\
\tableline
r700\_2 & 700 & 14.7 & a & 1.820 & 0.767 & 0.811 & 0.036 & 1.846 \\
        &     &      & b & 1.107 & 1.640 & 1.346 & 0.032 & 3.064 \\
\tableline
r700\_3 & 700 & 14.7 & a & 2.755 & 0.944 & 0.931 & 0.217 & 4.391 \\
\tableline
r600\_1 & 600 & 12.6 & a & 0.565 & 2.585 & 1.831 & 0.181 & 3.979 \\
        &     &      & b & 0.578 & 0.628 & 0.710 & 0.242 & 6.927 \\
        &     &      & c & 0.073 & 0.091 & 0.196 & 0.286 & 7.590 \\
        &     &      & d & 1.771 & 1.189 & 1.086 & 0.031 & 3.124 \\
\tableline
r400\_1 & 400 & 8.4  & a & 1.549 & 0.981 & 0.956 & 0.095 & 4.777 \\
        &     &      & b & 0.049 & 0.388 & 0.515 & 0.345 & 15.378 \\
\tableline
\end{tabular}
\end{center}
\end{table}

\clearpage

\begin{figure}\label{growth}
\includegraphics[scale=0.7]{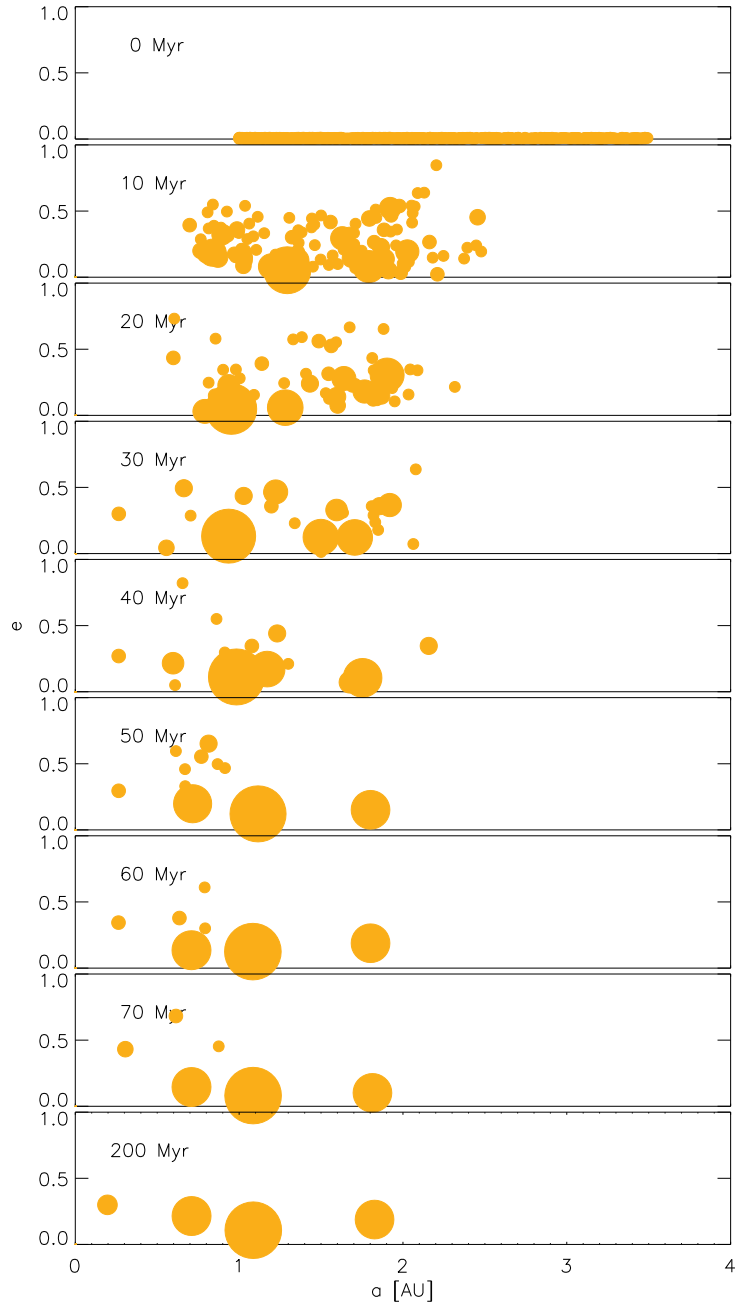}
\caption{Evolution of a circumstellar disk initially populated by lunar-mass
planetary embryos in nearly circular orbits around $\alpha$ Centauri B.  The 
radius of each circle is proportional to the size of the object.}
\end{figure}

\begin{figure}
\includegraphics{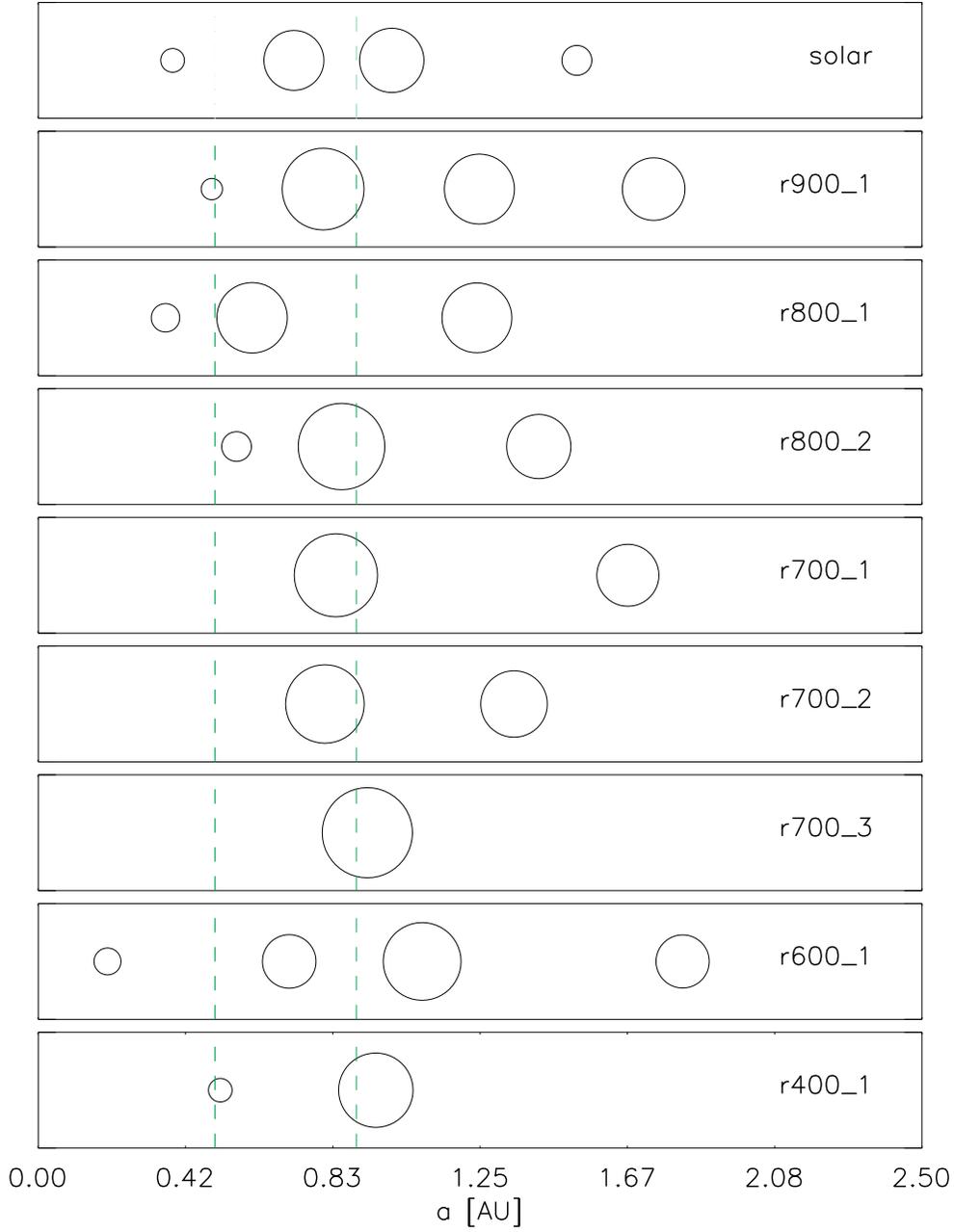}
\caption{Results from our planet formation simulations.  All simulations yield 
1-4 planets of which 42\% lie inside the star's habitable zone (dashed line). 
The planetary configuration of the solar system is shown for reference.}
\end{figure}

\begin{figure}
\includegraphics{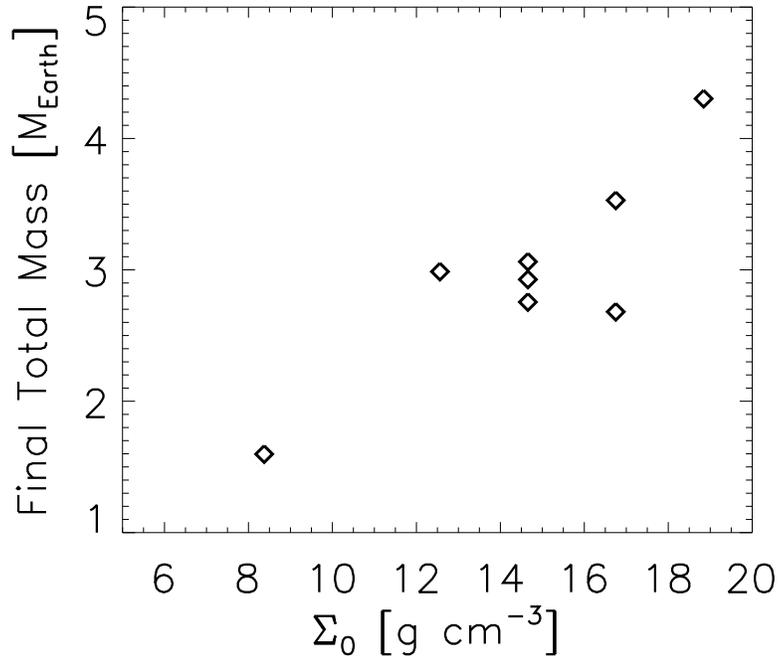}
\caption{Final mass of the resulting planetary systems as a function of initial 
surface density of the disk.}
\end{figure}

\begin{figure}
\includegraphics{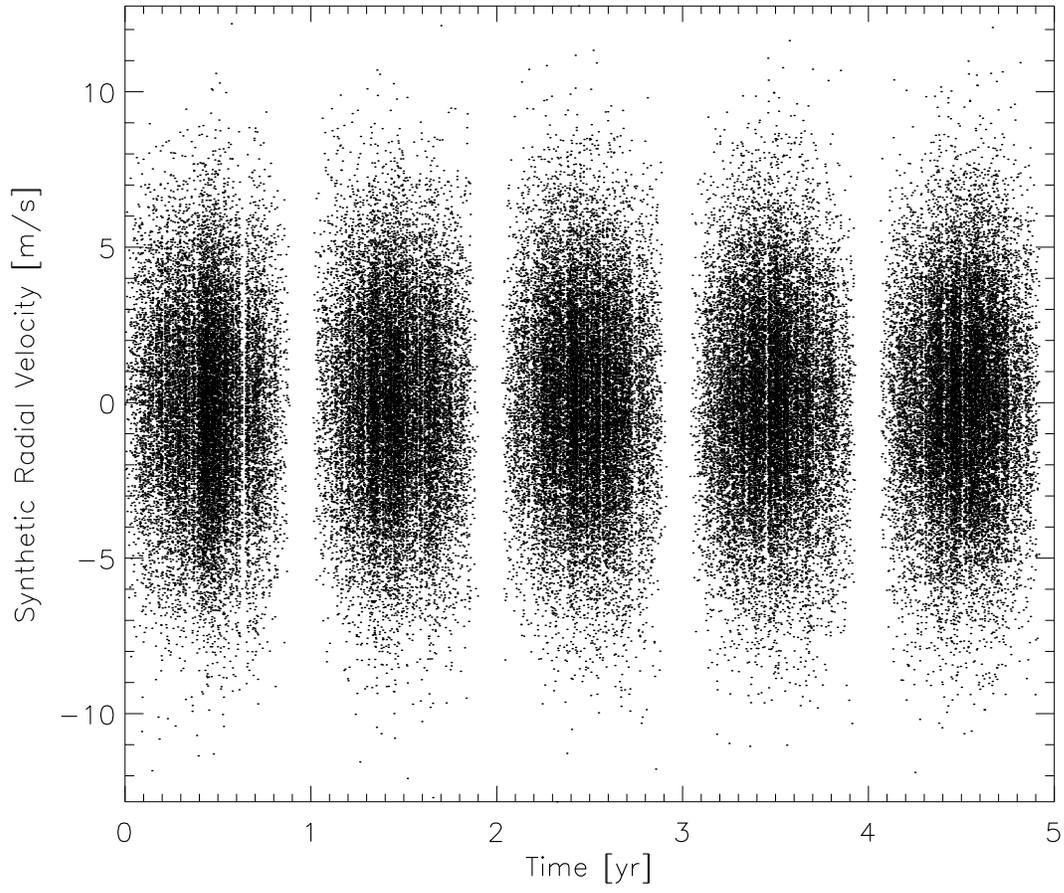}
\caption{Synthetic radial velocities over a five year period.  The 60-day gaps
in the data account for the time period when the system is below the horizon.
Other gaps in the data emulate nights of missed observations due to bad weather
and other adverse events.}
\end{figure}

\begin{figure}
\epsscale{0.7}
\includegraphics[scale=0.75]{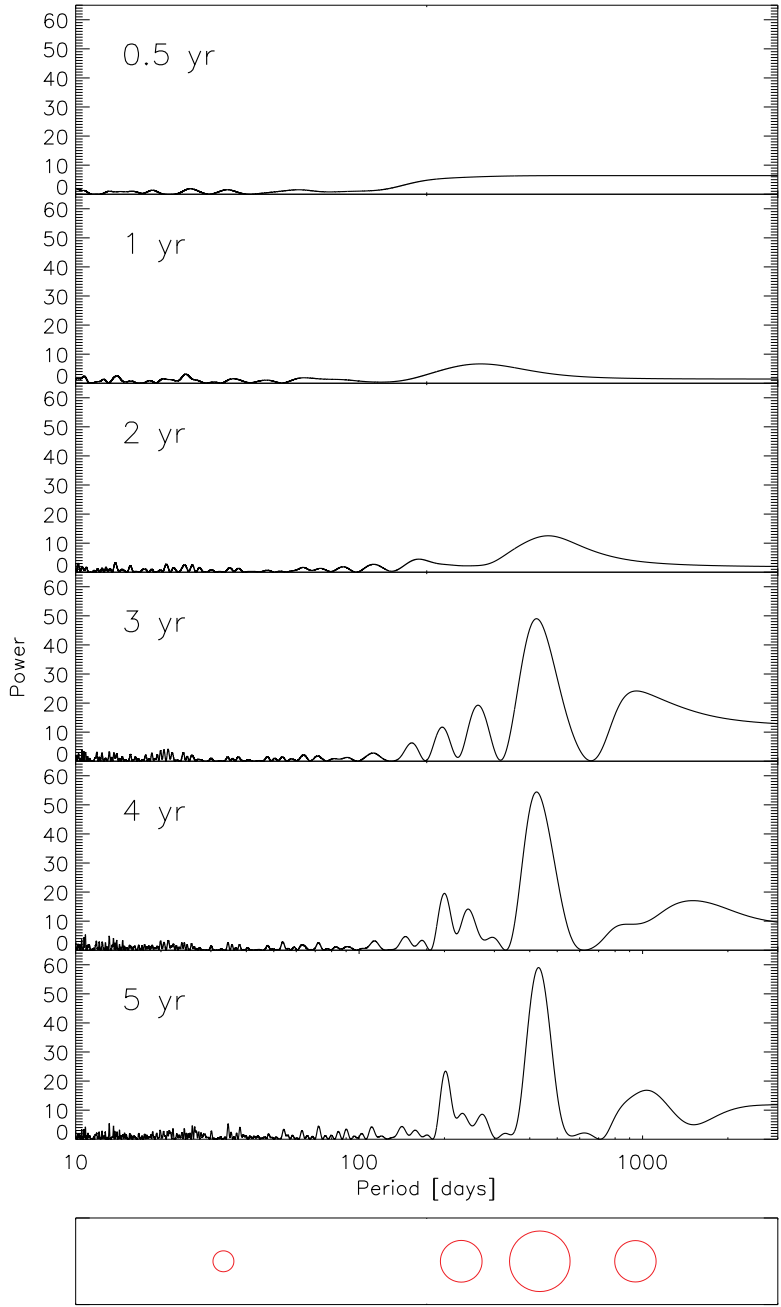}
\caption{Evolution of the periodogram for r600\_1 over 5 years as 97,260
synthetic radial velocity observations are made, assuming Gaussian white noise
with amplitude 3 m\,s$^{-1}$.  The 1.7 $M_{\oplus}$ planet (P = 1.2 yr) could be
confidently detected in 3 years.}
\end{figure}

\end{document}